\documentclass[paper=a4, 12pt, twoside]{article} 
\usepackage{lastpage}
\newcommand{\stpage}{1}  

\usepackage[top=2.54cm, bottom=2.54cm, left=2.54cm, right=2.54cm]{geometry} 
\usepackage{times} 
\usepackage{lmodern} 
\usepackage{hanging} 
\usepackage{graphicx} 
\usepackage[english]{babel}  
\usepackage{natbib}
\usepackage{fancyhdr} 
\fancypagestyle{frontpagefooter}{%
  \fancyhead{} 
  \fancyfoot[L]{\footnotesize 
    Corresponding Author: Anil K. Bera \\
    Email: abera@illinois.edu}
  \fancyfoot[C]{\footnotesize }	  
  \fancyfoot[R]{\footnotesize } 
}
\usepackage{amsthm}

\newtheoremstyle{sa}
{3mm} 
{2mm} 
{} 
{} 
{\bfseries} 
{:} 
{.5em} 
{} 

\theoremstyle{sa}

\usepackage{amsmath,amssymb}
\usepackage{hyperref} 
\usepackage{lscape}
\usepackage{arydshln}
\usepackage{booktabs}
\usepackage{verbatim}
\usepackage{enumitem}
\usepackage{bm}
\DeclareMathAlphabet{\mathpzc}{OT1}{pzc}{m}{it}
\usepackage{wasysym}
\usepackage{esint}
\usepackage{xcolor}
\usepackage{tikz}
\usetikzlibrary{shapes,backgrounds,arrows,chains,matrix,positioning,scopes}
\usepackage{caption}
\usepackage{titlesec}

\titleformat{\section}{\bf}
{\makebox[1.27cm][l]{\thesection.}}{0in}{} 
\titlespacing*{\section}{0pt}{3mm}{0pt}

\titleformat{\subsection}{\bf}
{\makebox[1.27cm][l]{\thesubsection.}}{0in}{}  
\titlespacing*{\subsection}{0pt}{3mm}{0pt}

\titleformat{\subsubsection}{\bf}
{\makebox[1.27cm][l]{\thesubsubsection.}}{0in}{}  
\titlespacing*{\subsubsection}{0pt}{3mm}{0pt}

\usepackage{indentfirst}  
\usepackage[labelfont=bf,textfont=bf]{caption}  
\usepackage{lineno}

\begin{document}
\thispagestyle{empty}

\thispagestyle{frontpagefooter}
 
%

%
%

\begin{center}
{\fontsize{16}{24} \selectfont 
{\bf Three Score and 15 Years (1948-2023) of Rao's Score Test: A Brief History 
	} 
 } \\[6mm]
\normalsize
{\bf $^1$Anil K. Bera and $^2$Yannis Bilias} \\ 
{\it $^1$University of Illinois at Urbana-Champaign, U.S.A.\\
$^2$Athens University of Economics and Business, Greece}\\

Received: 15 June 2024; Revised: 03 September 2024; 
\end{center}
\vspace{-6mm}

\setlength\parindent{1.27cm}

\noindent\rule{16.5cm}{0.8pt}\vspace{-3mm}
\\
\begin{figure}[h]
	\centering
	\includegraphics[width=0.8\textwidth, height=0.1\textheight]{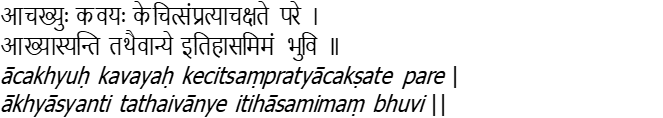}
	
	\hfill Mahabharata ($\sim$500BC) by Krishna-Dwaipayana Vyasa,
	Verse 1.1.24
\end{figure}
\begin{center}
\vspace*{-10mm}
``\textsf{Some bards have already published this history, 
some are now teaching it, \\
and others, in like manner, will
hereafter promulgate it upon the earth}."

\end{center}

\vspace*{-1mm}

\noindent\rule{16.5cm}{0.8pt}\vspace{-3mm}

\noindent{\bf Abstract}\vspace{-1mm}

\cite{Rao:1948} introduced the score test statistic as an alternative to the likelihood ratio and Wald test statistics. In spite of the optimality properties of the score statistic shown in \cite{RaoPoti:1946}, the Rao score (RS) test remained unnoticed for almost 20 years. 
Today, the RS test is part of the ``Holy Trinity" of hypothesis testing and has found its place 
in Statistics and Econometrics textbooks and related software.  
Reviewing the history of the RS test we note that remarkable test statistics proposed in 
the literature earlier or around the time of \cite{Rao:1948} mostly from intuition, 
such as \cite{Pearson:1900} goodness-fit-test, \cite{Moran:1948} \textbf{I} test for spatial dependence and \cite{DW1950} test for serial correlation, can be given RS test statistic interpretation.  At the same time, recent developments in the robust hypothesis testing under certain forms of misspecification, make the RS test
an active area of research in Statistics and Econometrics. From our brief account of the history of the RS test we conclude that its impact in science goes far beyond its calendar starting point with promising future research activities for many years to come.

\setcounter{equation}{0}

\\

\vspace{-6mm}

\setlength\parindent{1.27cm}

\noindent\rule{16.5cm}{0.8pt}\vspace{-3mm}

\section{Prologue}

C. R. Rao’s work was always inspired by some practical problems. In 1946, he was deputed 
from the Indian Statistical Institute (ISI), Calcutta,  to work on an anthropometric project in the Museum of Anthropology and Ethnology at the Cambridge University, U.K. While at Cambridge, Rao took the opportunity to contact R. A. Fisher, then the Belfour Professor of Genetics, and registered for a Ph.D. degree in Statistics under Fisher’s guidance. As recollected in 
\cite{Rao:2001}, Fisher agreed under the condition that Rao spends time in the
Genetics Laboratory where Fisher was breeding mice to map their chromosomes.
Rao started by mating mice of different genotypes to collect the necessary data 
and additionally, he was trying to develop appropriate statistical methodology to analyze the experimental data. The problem was estimation of linkage \textit{parameters} (recombination probabilities in the various segments of the chromosomes) using data sets from different experiments, designed in such a way that each data set had information on the \textit{same parameters}. It was thus necessary to test whether the parameters in different experiments are the same or not.

Rao wrote and published two papers based on this work. The \textit{first} paper, \cite{Rao:1948}, deals with the \textit{general problem of testing} simple and composite hypotheses concerning a vector parameter. The test was based on the \textit{scores}, derivatives of the log-likelihood function with respect to the individual parameters. The paper was published in the \textit{Proceedings of the Cambridge Philosophical Society},  
where he termed the test principle as a 
\textit{score test}. In this paper, we will refer to it as the Rao score (RS) test. The \textit{other} paper, \cite{Rao:1950}, contains the detail steps for \textit{analyzing the data} involving the segregation of several factors in mating of different genotypes. And it used the RS test for the meta-analysis of testing the equality of parameters coming from different experimental data sets. That paper was published in Fisher’s new journal \textit{Heredity}. 
For more, see \cite{Rao:2001}.

The rest of the paper is organized as follows.
In Section 2, we start with the first principle of testing, namely the Neyman-Pearson Lemma and derive the simplest version of RS test and then discuss it in its full generality. 
There, we also provide RS test interpretation to some of the classic tests in Econometrics and Statistics, such as the quintessential \cite{Pearson:1900} goodness-fit-test, 
which was 
suggested mostly by pure intuition, but its theoretical foundation can 
be buttressed by RS test principle. 
In Sections 3 and 4, we list a (somewhat incomplete) catalogue of 
RS tests in Econometrics and Statistics. 
In Section 5, we outline some of the possible ways an assumed probability model can be 
misspecified, and discuss how the various RS tests can be robustified to make them valid
under misspecification. We close the paper in Section 6 (Epilogue) with some concluding 
remarks. 
At the outset let us mention that while compiling the 75 years 
(from 1948 to 2023)
history of the RS test, we have included here some of our own past historical accounts 
and cited accordingly. 
Our aim is to have a comprehensive review as far as possible at one place, like a 
one-stop-shopping for the RS test history.

\section{Score as an optimal test function: \cite{RaoPoti:1946}}
We start by introducing some notation and concepts. Suppose we have $n$ independent observations $y_1, y_2,..., y_n$ on a random variable $Y$ with density function $f(y; \theta)$, where $\theta$ is a $p\times 1$ parameter vector with $\theta \in \Theta \subseteq \mathbb{R}^p$. It is assumed that $f(y; \theta)$ satisfies the regularity conditions stated in 
\citet[p.364]{Rao:1973} and \citet[p.144]{Serfling:1980}. The likelihood function is given by
\begin{equation}
    \textbf{L}(\theta, \textbf{y}) \equiv \textbf{L}(\theta) = \prod_{i = 1}^{n} f(y_i; \theta),
\end{equation}
where \textbf{y} $= (y_1, y_2, \ldots , y_n)$ denotes the sample. Suppose we want to test $H_0: \theta = \theta_0$ against $H_1: \theta \neq \theta_0$ based on the sample \textbf{y}. 

The foundation of the theory of hypothesis testing was laid by
\cite{NeymanPearson:1933} fundamental lemma.
This lemma provides a way to find the most powerful (MP) and uniformly most powerful (UMP) tests. According to the Neyman-Pearson (N-P) Lemma, the MP critical region
for testing $H_0 : \theta = \theta_0$ versus $H_1 : \theta = \theta_1$ having size $\alpha$, is given by
\begin{equation}
    \omega(\textbf{y}) = \{\textbf{y} \; | \; \textbf{L}(\theta_1) > \kappa\textbf{L}(\theta_0)\}, 
\end{equation}
where $\kappa$ is such that $\Pr[{\omega(\textbf{y}) | H_0}] = \alpha$. 

If an MP test maximizes powers uniformly in $\theta_1 \in \Theta_1 \subseteq \Theta$, the test is called UMP test. Unfortunately, an UMP test rarely exists, and when it does not, there is no single critical region best for all alternatives. We, therefore, try to find a critical region that is good for alternatives \textit{close} to the null hypothesis, called \textit{local} alternatives, hoping that the region will also be good for alternatives away from the null. \citet[p.529]{Lehmann:1999} advocated for such critical region when the sample size $n$ is large, stating, ``if the true value is at some distance from $\theta_0$, a large sample will typically reveal this so strikingly that a formal test may be deemed unnecessary." 

For a critical region $\omega(\textbf{y})$, let us define the power function as
\begin{equation}
    \gamma(\theta) = \Pr[\omega(\textbf{y}) | \theta ] = \int_{\omega(\textbf{y})} \textbf{L}(\theta)d\textbf{y}.
\end{equation}
Assuming a scalar $\theta$ and that $\gamma$($\theta$) admits Taylor series expansion, we have  
\begin{equation}
    \gamma(\theta) = \gamma(\theta_0) + (\theta-\theta_0)\gamma^\prime(\theta_0) + \frac{(\theta - \theta_0)^2}{2}\gamma^{\prime \prime}(\theta^*),
\end{equation}
where $\theta^*$ is a value in between $\theta$ and $\theta_0$. If we consider local alternatives of the form $\theta = \theta_0 + n^{-\frac{1}{2}}\delta, 0 < \delta < \infty$, the third term will be of the order \textbf{O}($n^{-1}$). Therefore, from (4), to obtain the highest power, we need to maximize
\begin{equation}
    \gamma^\prime(\theta_0) = \frac{\partial}{\partial \theta} \gamma(\theta) \Biggr|_{\theta = \theta_0} = \frac{\partial}{\partial \theta} \int_{\omega(\textbf{y})} \textbf{L}(\theta_0)d\textbf{y}
    = \int_{\omega(\textbf{y})} \frac{\partial}{\partial \theta} \textbf{L}(\theta_0)d\textbf{y},
\end{equation}
for $\theta > \theta_0$, assuming the regularity conditions that allow differentiation under the sign of intergration. 

Using the generalized N-P Lemma given in \cite{NeymanPearson:1936}, it is easy to see that the locally most powerful (LMP) critical region for $H_0 : \theta = \theta_0$ versus $H_1 : \theta > \theta_0$, is given by
\begin{align*}
     \omega(\textbf{y}) = \left\{\textbf{y} \;\Big| \; \frac{\partial}{\partial \theta} \mathbf{L}(\theta_0) > \kappa\mathbf{L}(\theta_0) \right\}
\end{align*}
or  
\begin{equation}
	 \omega(\textbf{y}) = \left\{\textbf{y} \; \Big| \;
\frac{\partial}{\partial \theta} \ln(\mathbf{L}(\theta_0)) = \frac{\partial}{\partial \theta} l(\theta_0) > \kappa \right\},
\end{equation}
where $l(\theta)$ denotes the log-likelihood function and $\kappa$ is a constant such that the size of the test is $\alpha$. The quantity $S(\theta) = {\partial l(\theta)}/{\partial \theta}$ is called the Fisher-Rao \textit{score function}. The above result in (6) was first discussed in \cite{RaoPoti:1946}, who stated that a LMP critical region for $H_0 : \theta = \theta_0$ is given by
\begin{equation}
   \omega(\textbf{y}) = \{\textbf{y} \; | \; \kappa_1 S(\theta_0) > \kappa_2 \},
\end{equation}
where $\kappa_2$ is so determined that the size of the test is equal to a preassigned value $\alpha$ with $\kappa_1$ as $+1$ or $-1$, respectively, for alternatives $\theta > \theta_0$ and $\theta < \theta_0$. 

Let us define the Fisher information as
\begin{equation}
    \mathcal{I}(\theta) = -E \left[\frac{\partial^2 l(\theta)}{\partial\theta^2}\right] = Var[S(\theta)]. 
\end{equation}
The result that under $H_0$, $S(\theta_0)$ is asymptotically distributed as normal with mean zero and variance $\mathcal{I}(\theta_0)$, led \cite{RaoPoti:1946} to suggest a test based on $S(\theta_0)/\sqrt{\mathcal{I}(\theta_0)}$ as standard normal [or $S^2(\theta_0)/\mathcal{I}(\theta_0)$ as $\chi^2_1$], for large $n$.

\subsection{From \cite{RaoPoti:1946} to \cite{Rao:1948}: Test for the multiparameter\\\hspace*{-1.27cm}case}

\cite{RaoPoti:1946} can be viewed as a precursor to \cite{Rao:1948}. Generalization of the 
LMP test in (7) to the multiparameter case ($p \ge 2$) is not trivial. There will be scores for each individual paramter, and the problem is to combine them in an ``optimal" way. Let $H_0: \theta = \theta_0$, where now $\theta = (\theta_1, \theta_2,...,\theta_p)^\prime$ and $\theta_0 = (\theta_{10}, \theta_{20},...,\theta_{p0})^\prime$. 
Consider a $\textit{scalar}$ linear combination 
\begin{equation}
    \sum_{j=1}^{p} \delta_j \frac{\partial l(\theta)}{\partial \theta_j} = \delta^\prime S(\theta),
\end{equation}
where $\delta = (\delta_1, \delta_2,..., \delta_p)^\prime$ is a fixed vector and test the hypothesis $H_{0 \delta}$ : $\delta^\prime \theta = \delta^\prime \theta_0$ against $H_{1 \delta}$ : $\delta^\prime \theta \neq \delta^\prime \theta_0$, $\delta \in \mathbb{R}^p$.

We rewrite the Fisher information in (8) as
\begin{equation}
    \mathcal{I}(\theta) = -E \left[\frac{\partial^2 l(\theta)}{\partial \theta \partial \theta^\prime}\right].
\end{equation}
Asymptotically, under $H_0$, $\delta^\prime S(\theta_0)$ is distributed as normal, with mean zero and variance $\delta^\prime \mathcal{I}(\theta_0)\delta$. Thus if $\delta$'s were known, a test could be based on
\begin{equation}
    \frac{\left[\delta^\prime S(\theta_0)\right]^2}{\delta^\prime \mathcal{I}(\theta_0)\delta},
\end{equation}
which under $H_0$ will be distributed as $\chi^2_1$ as in \cite{RaoPoti:1946}. Note that our $H_0 : \theta = \theta_0$ for $p \ge 2$ can be expressed as $H_0 \equiv \underset{\delta \in \mathbb{R}^p}{\bigcap} H_{0\delta}$, i.e., the multiparameter testing problem can be decomposed into a series of \textit{single-parameter} problems. To obtain a linear function like (9), \cite{Rao:1948} maximized (11) with respect to $\delta$. Using the Cauchy-Schwarz inequality
\begin{equation}
    \frac{(u^\prime v)^2}{u^\prime Au} \leq v^\prime A^{-1}v,
\end{equation}
where $u$ and $v$ are column vectors and $A$ is a non-singular matrix, we have
\begin{equation}
    \underset{\delta \in \mathbb{R}^p}{\sup} \frac{\left[\delta^\prime S(\theta_0)\right]^2}{\delta^\prime \mathcal{I}(\theta_0)\delta} = S(\theta_0)^\prime \mathcal{I}(\theta_0)^{-1} S(\theta_0).
\end{equation}
In (13), the supremum reaches at $\delta = \mathcal{I}(\theta_0)^{-1} S(\theta_0)$ and this provides an optimal linear combination of scores. 

\cite{Roy1953} used Rao's maximization technique (13) to develop his \textit{union-intersection} (UI) method of testing. Let $H_0 \equiv \underset{j \in J}{\bigcap} H_{0j}$, where $J$ is an index set. Roy's UI method gives the rejection region for $H_0$ as the \textit{union} of rejection regions for all $H_{0j}$, $j \in J$. Consider testing $H_{0\delta}:$ $\delta^\prime \theta = \delta^\prime \theta_0$ against $H_{1\delta}:$ $\delta^\prime \theta \neq \delta^\prime \theta_0$, $\delta \in \mathbb{R}^{p}$. Let $H_0 = \underset{\delta \in \mathbb{R}^p}{\bigcap} H_{0\delta}$ and $H_1 \equiv \underset{\delta \in \mathbb{R}^p}{\bigcap} H_{1\delta}$. If $T_\delta$ is the likelihood ratio (LR) statistic for testing $H_{0\delta}$ against $H_{1\delta}$, then 
\begin{equation}
    T = \underset{\delta \in \mathbb{R}^p}{\sup} T_\delta
\end{equation}
is Roy's LR statistic for testing $H_0$ against $H_1$. This is the same principle that was 
used by \cite{Rao:1948} to convert a ``multivariate" problem into a series of ``univariate" ones, as we have seen in equation (13).

When the null hypothesis is composite, like $H_0:h(\theta) = c$, where $h(\theta)$ is an $r\times 1$ vector function of $\theta$ with $r \leq p$, the general form of the RS test statistic is
\begin{equation}
    RS = S(\widetilde{\theta})^\prime \mathcal{I}(\widetilde{\theta})^{-1} S(\widetilde{\theta}), 
\end{equation}
where $\widetilde{\theta}$ is the restricted maximum likelihood estimator (MLE) of $\theta$, i.e., $h(\tilde{\theta}) = c$.   Asymptotically, under $H_0$, the RS test statistic is distributed as {$\chi^{2}_{r}$}. Therefore, we observe \textit{two} optimality principles behind the RS test; first, in terms of LMP test as given in (6), and second, in deriving the ``optimal" direction for the multiparameter case, as in (13). 

\cite{Rao:1948} suggested the score test as an alternative to the \cite{Wald:1943} statistic, which for testing $H_0: h(\theta) = c$ is given by
\begin{equation}
    W = \left[h(\hat{\theta}) - c\right]^\prime \left[H(\hat{\theta})^\prime \mathcal{I}(\hat{\theta})^{-1} H(\hat{\theta})\right]^{-1} \left[h(\hat{\theta}) - c\right],
\end{equation}
where $\hat{\theta}$ is the unrestricted MLE of $\theta$, and $H(\theta) = {\partial h(\theta)}/{\partial \theta}$ is a $r\times p$ matrix with full column rank $r$. \citet[p.53]{Rao:1948} stated that his test (15), ``besides being simpler than Wald's has some theoretical advantages." For more on this see \cite{Bera:2000} and \cite{Bera:2001}.

\cite{NeymanPearson:1928} suggested their LR test as 
\begin{equation}
    LR = 2 \left[ln \frac{\textbf{L}(\hat{\theta})}{\textbf{L}(\widetilde{\theta})}\right] = 2 \left[l(\hat{\theta}) - l(\widetilde{\theta})\right].
\end{equation}
Their suggestion did not come from any search procedure satisfying an optimality criterion. It was purely based on intuitive grounds; as \citet[p.6]{Neyman:1980} stated, ``The intuitive background of the likelihood ratio test was simply as follows: if among the contemplated admissible hypotheses there are some that ascribe to the facts observed probabilities much larger than that ascribed by the hypothesis tested, then it appears `reasonable' to reject the null hypothesis."

The three statistics RS, W, and LR, given respectively in (15), (16), and (17) are referred to as the ``Holy Trinity." These tests can be viewed as three different \textit{distance measures} between $H_0$ and $H_1$. When $H_0$ is true, we should expect the restricted and unrestricted MLEs of $\theta$, namely $\widetilde{\theta}$ and $\hat{\theta}$ to be close, and likewise the log-likelihood functions $l(\widetilde{\theta})$ and $l(\hat{\theta})$, respectively. The LR statistic in (17) measures the distance through the log-likelihood function and is based on the difference $l(\hat{\theta}) - l(\widetilde{\theta})$. To see the intuition behind the RS test, note that $S(\hat{\theta}) = 0$ by construction, and thus we should expect $S(\widetilde{\theta})$ to be cloes to zero if $H_0$ is true. Therefore, the basis of the RS test is $S(\widetilde{\theta}) - S(\hat{\theta}) = S(\widetilde{\theta})$,  distance between $\widetilde{\theta}$ and $\hat{\theta}$ measured through the 
function $S(\theta)$. Finally to test $H_0: h(\theta) = c$, W considers the distance directly in terms of $h(\theta)$, and is based on $\left[h(\hat{\theta}) - c\right] - \left[h(\widetilde{\theta}) - c\right] = h(\hat{\theta}) - c$, where $h(\widetilde{\theta}) = c$ by construction, as we see in expression (16). It is interesting to note the similarity between the Wald and the RS tests based on $h(\hat{\theta})$ and $S(\widetilde{\theta})$, respectively. Therefore the RS test statistic is closer to W than LR. Therefore, it makes sense that 
\citet[p.53]{Rao:1948}
mentioned his test as an alternative to W.

The interrelationships among these three tests can be brought home to the students of Statistics through the following amusing story [see \citet[p.58]{Bera:chap2001}]:
Once around 1946 Ronald Fisher invited Jerzy Neyman, Abraham
Wald, and C.R. Rao to his  Cambridge University lodge 
for afternoon tea. During their conversation, Fisher
mentioned the problem of deciding whether his dog, who had been going to an 
“obedience school” for some time, was disciplined enough. Neyman quickly came up with
an idea: leave the dog free for some time and then put him on leash. If there is not
much difference in his behavior, the dog can be thought of as having completed the
course successfully. Wald, who lost his family in the concentration camps, was adverse
to any kind of restrictions and simply suggested leaving the dog free and seeing whether it
behaved properly. Rao, who had observed the nuisances of stray dogs in Calcutta
streets, did not like the idea of letting the dog roam freely, and suggested keeping the
dog on a leash at all times and observing how hard it pulls on the leash. If it pulled
too much, it needed more training. That night when Rao was back in his Cambridge
dormitory after tending Fisher’s mice at the genetics laboratory, he suddenly realized
the connection of Neyman and Wald’s recommendations to the Neyman–Pearson LR
and Wald tests, respectively. He got an idea and the rest, as they say, is history.

At this stage, it will be instructive to provide a geometric illustration highlighting the fundamental connections and contrasts among the three tests [see \citet[pp.56-60]{Bera:1983Phd}]. For simplicity, let us 
consider the case of scalar $\theta$, i.e., $p = 1$, and that the null hypothesis is $H_0: \theta = \theta_0$. In Figure 1, we plot the score function $S(\theta) = {d l(\theta)}/{d \theta}$ against $\theta$, the solid curved line. The unrestricted MLE $\hat{\theta}$ is obtained by setting $S(\hat{\theta}) = 0$, i.e., at the point D. From the Figure 1, it is easily seen that 
\begin{align}
\nonumber    l(\hat{\theta}) - l(\theta_0) & = \int_{\theta_0}^{\hat{\theta}}S(\theta)d(\theta) = \\
     &= \mbox{Area under the curve } S(\theta) \mbox{ from } \theta_0 \mbox{(point C) to } \hat{\theta} \mbox{(point D)}.
\end{align}
Therefore, 
\begin{equation}
    LR = 2[l(\hat{\theta}) - l(\theta_0)] = 2\cdot Area(CDF).
\end{equation}
For our particular case, $h(\theta) = \theta - \theta_0, H(\theta) = 1$ and $c = 0$. Thus, W in (16) can be expressed as 
\begin{equation}
    W = (\hat{\theta} - \theta_0)^2\mathcal{I}(\hat{\theta}) = CD^2 \cdot \mathcal{I}(\hat{\theta}).
\end{equation}
$\mathcal{I}(\hat{\theta})$ can be obtained from $-{d^2l(\theta)}/{d\theta^2} = -{dS(\theta)}/{d\theta}$, evaluated at $\theta = \hat{\theta}$, i.e., from $\tan\phi_{\hat{\theta}} = {CG}/{CD}$. Therefore, 
\begin{equation}
    W = CD^2 \cdot \frac{CG}{CD} = CD \cdot CG = 2\cdot Area(\bigtriangleup CDG).
\end{equation}
On the other hand, the RS test will be based on $S(\theta)$ at $\theta_0$, i.e., on the distance CF. 
\begin{figure}
    \centering
    \includegraphics{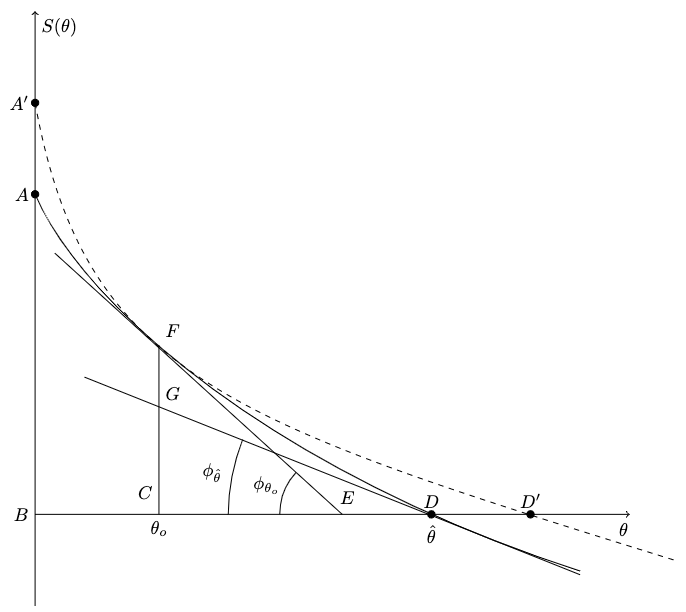}
    \caption{Geometry of LR, W and RS statistics}
\end{figure}
The variance of $S(\theta_0)$ can be estimated by $-{dS(\theta_0)}/{d\theta} = 
\tan\phi_{\theta_0} = {CF}/{CE}$. Hence, 
\begin{equation}
    RS = CF^2.\quad \frac{CE}{CF} = CF \cdot CE = 2\cdot Area(\bigtriangleup CEF).
\end{equation}

From above, we can note the following features of the three tests. \textit{First}, since the tests are based on three different areas in general, they will yield conflicting inference if the same critical value is used [see \cite{Savin:1977}]. \textit{Second}, the RS tests depends only on $S(\theta)$ and the slope of $S(\theta)$ at $\theta_0$. We can draw many curved lines through F with the same slope at F, and the dotted line $A^\prime FD^\prime$ is an example. This implies that there may be other $S(\theta)$ functions, i.e., other likelihood function representing different alternative hypothesis, with the same slope at $\theta_0$, giving rise to the same RS test statistic. In the literature, this property is known as invariance property of the RS test principle [see \citet[p.70]{Godfrey:1988}]. \textit{Finally}, for both the RS and W tests, the variances can be calculated in a number of ways which are asymptotically equivalent. This can lead to different versions of the test statistics. It is not clear which versions will give better results in finite samples. 

\textbf{Example 1:} Let us start with a simple example where $y_i \sim IID N(\theta,1), \ i = 1, 2,\ldots, n$, and we test $H_0: \theta = \theta_0=0$ against $\theta > 0$. Here the log-likelihood and 
score functions are respectively 
\begin{equation}
    \begin{split}
        l(\theta) &= \text{Constant}-\frac{1}{2}\sum^n_{i=1}(y_i-\theta)^2, \\
        \text{and} \quad S(\theta) &= \sum_{i = 1}^n(y_i - \theta)=n(\bar{y}-\theta),
    \end{split}
\end{equation}
where $\bar{y}=\sum_i y_i/n$. 
Note that here $S(\theta)$ is \textit{linear} in $\theta$, and thus from Figure 1, all the three tests LR, W, and RS will be identical. Given that $S(\theta_0) = n\bar{y}$ with $Var[S(\theta_0)] = n$,  we will reject $H_0$, if $\sqrt{n}\Bar{y} > \textbf{z}_\alpha$, where $\textbf{z}_\alpha$ is the upper $\alpha$ percent cut-off point of standard normal 
distribution. For fixed $n$, the power of this test goes to 1 as $\theta \rightarrow \infty$. Hence the score test $\sqrt{n}\Bar{y} > \textbf{z}_\alpha$ is not only LMP, but also UMP for all $\theta > 0$.

\textbf{Example 2:} [\citet[p.235]{Ferguson:1967}] Consider testing for the median of a Cauchy distribution with density 
\begin{equation}
    f(y;\theta) = \frac{1}{\pi}\cdot\frac{1}{1 + (y-\theta)^2}, \quad -\infty < y < \infty.
\end{equation}
Since here $\mathcal{I}(\theta)=n/2$, 
the RS test will reject $H_0: \theta = \theta_0$ against $H_1: \theta > \theta_0$, if
\begin{equation}
    \frac{S(\theta_0)}{\sqrt{\mathcal{I}(\theta_0)}} = \sqrt{\frac{2}{n}}\sum^{n}_{i = 1} \frac{2(y_i-\theta_0)}{1+(y_i-\theta_0)^2} > \textbf{z}_\alpha.
\end{equation}
As $\theta \rightarrow \infty$ with $n$ remaining fixed, $\min(y_i - \theta_0) 
\overset{p}{\rightarrow} \infty$, and ${S(\theta_0)}/{\sqrt{\mathcal{I}(\theta_0)}}
\overset{p}{\rightarrow} 0$. Thus the power of the test tends to zero
as $\theta \rightarrow \infty$. Therefore what works for \textit{local} alternatives may not work for \textit{not-so-local} alternatives. This is in contrast to Example 1 where the LMP test is also the UMP. 

In the example below we illustrate one of the most famous tests in the Statistics literature that was suggested long before 1948 and the theoretical foundation of which 
can be buttressed by the RS test principle. 

\textbf{Example 3:} [\cite{Pearson:1900} Goodness-of-fit test]. Consider a multinomial distribution with $p$ classes and let the probability of an observation belonging to the $j$-th class be $\theta_j (\ge 0)$, $j = 1, 2, \ldots, p$, so that $\sum^p_{i = 1}\theta_j = 1$. Denote the observed frequency of the $j$-th class by $n_j$ with $\sum^p_{j = 1} n_j = n$. We are interested in testing $\theta_j = \theta_{j0}$, $j = 1,2, \ldots, p$, where $\theta_{j0}$ are known constants. \cite{Pearson:1900} suggested the statistic
\begin{equation}
    P = \sum^p_{j = 1}\frac{(n_j - n\theta_{j0})^2}{n\theta_{j0}} = \sum\frac{(O - E)^2}{E},
\end{equation}
where $O$ and $E$ denote respectively, the observed and expected frequencies. Given the profound importance of $P$ in almost all branches of science, we demonstrate the theoretical 
underpinnings of $P$ based on the RS test principle. The log-likelihood function, score and information matrix are respectively, given by [see \citet[p.17]{Bera:2001}]
\begin{equation}
    l(\theta) = Constant + \sum^p_{j = 1}n_j \ln(\theta_j)
\end{equation}
\begin{equation}
    \underset{{[(p-1) \times 1]}}{S(\theta)} = 
    \begin{bmatrix}
        \frac{n_1}{\theta_1} -\frac{n_p}{\theta_p} \\
        \frac{n_2}{\theta_2}  -\frac{n_p}{\theta_p} \\
        \cdots \\
        \frac{n_{p-1}}{\theta_{p-1}}  -\frac{n_p}{\theta_p}
    \end{bmatrix}
\end{equation}
and
\begin{equation}
    \underset{{[(p-1) \times (p-1)]}}{\mathcal{I}(\theta)} = n\left[diag\left(\frac{1}{\theta_1}, \frac{1}{\theta_2}, \ldots, \frac{1}{\theta_{p-1}}\right) + \frac{1}{\theta_p}\textbf{1}\textbf{1}^\prime\right]
\end{equation}
where \textbf{1} $= (1,  1, \ldots,1)^\prime$ is a $(p-1) \times 1$ vector of ones. We end up with effectively $(p-1)$ parameters since $\sum^p_{j=1}\theta_j = \sum^p_{j=1}\theta_{j0} = 1$. Using the above expressions, it is easy to see that 
\begin{equation}
    S(\theta_0)^\prime\mathcal{I}(\theta_0)^{-1}S(\theta_0) = P, 
\end{equation}
where $\theta_0 = (\theta_{10}, \theta_{20},..., \theta_{p0})^\prime$ [see 
\citet[p.442]{Rao:1973} and \cite[p.316]{CoxHinkley:1974}]. The coincidence that $P$ is same as the RS test, is an amazing result. \cite{Pearson:1900} suggested his test mostly based on intuitive grounds almost 50 years before \cite{Rao:1948}.
\section{Some applications of the RS test in Econometrics}

RS test was well ahead of its time. It went unnoticed for very many years. It is fair to say that econometricians can claim major credit in
recognizing its importance and
applying the RS test in several useful contexts and coming up with closed form, neat test statistics. 
Rao himself acknowledged this fact by 
writing [see \citet[p.15]{Rao2005}] ``I am gratified to see the large number of papers contributed by 
econometricians on the application of the score statistic to problems in econometrics 
and the extensions and improvements they have made." 
More recently, statisticians are catching up with innovative applications.
To obtain a quantitative perception of the influence of \cite{Rao:1948}, 
we plot the yearly citations
for the last 75 years in Figure 2. The corresponding cumulative citations are depicted 
in Figure 3. \textit{First} thing to note is that the total number of citations in the 
last 75 years is only 980, apparently a very low number for such a seminal paper. 
Of course, we need to take into consideration of the fact that there are many papers, 
especially in the Statistics literature, that use the score test without making any 
reference to \cite{Rao:1948}. 
\textit{Second}, there are only a handful citations during the first thirty years, i.e., 
until around 1978. That was the time econometricians recognized the usefulness of the Rao test
principle, and used it in developing several model specification tests.
There was another surge in its use after another 30 years, i.e., around 2008, in both
the Statistics and Econometrics literature. 
\textit{Finally}, from both Figures 2 and 3, it is clear that overall,
the number of citations is still 
going up at 
an increasing rate, indicating continuing influence of \cite{Rao:1948}, as far as the citation
numbers go.
\begin{figure}[h]
		\begin{minipage}[c]{0.45\linewidth}
			\includegraphics[width=\linewidth]{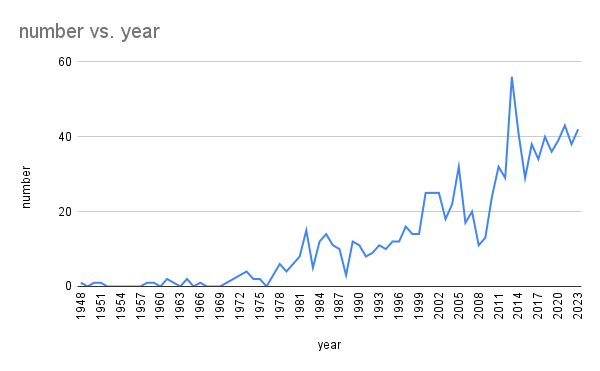}
			\caption{Yearly number of citations of Rao (1948): 1948-2023}
		\end{minipage}
		\hfill
		\begin{minipage}[c]{0.45\linewidth}
			\includegraphics[width=\linewidth]{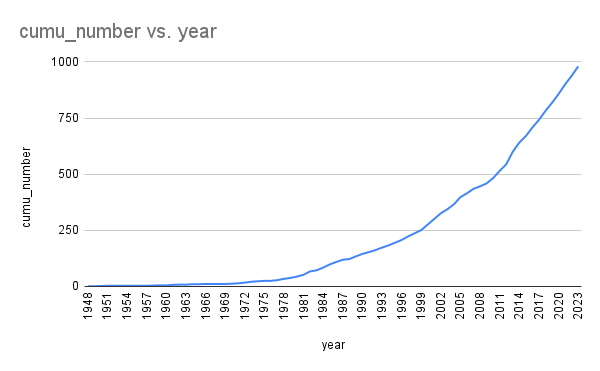}
			\caption{Cumulative number of citations }
		\end{minipage}%
\end{figure}

\cite{Byron:1968} was probably the first to apply the RS test in Econometrics. He used \cite{Silvey:1959} Lagrange multiplier (LM) version along with the LR statistic for testing homogeneity and symmetry restrictions in the demand system. In the Econometrics literature, the RS test is known as the LM test - the terminology came from \cite{Silvey:1959}. Note that the restricted MLE $\widetilde{\theta}$ under the restriction $H_0: h(\theta) = c$ can be obtained from the first order condition of the Lagrangian function
\begin{equation}
    \mathcal{L} = l(\theta) - \lambda^\prime[h(\theta) - c],
\end{equation}
where $\lambda$ is an $r \times 1$ vector of Lagrange multipliers. The first order conditions are
\begin{equation}
    S(\widetilde{\theta}) - H(\widetilde{\theta})\widetilde{\lambda} = 0
\end{equation}
\begin{equation}
    h(\widetilde{\theta}) = c,
\end{equation}
where $H(\theta) = {dh(\theta)}/{d\theta}$. Therefore, from (32) we have $S(\widetilde{\theta}) = H(\widetilde{\theta})\widetilde{\lambda}$. Given that $H(\theta)$ has full rank, $S(\widetilde{\theta})$ = 0 is equivalent to $\widetilde{\lambda} = 0$. These multipliers can be interpreted as the implicit cost (shadow prices) of imposing the restrictions $h(\theta) = c$. It can be shown that 
\begin{equation}
    \widetilde{\lambda} = \frac{dl(\widetilde{\theta})}{dc},
\end{equation}
i.e., the multipliers give the rate of change of the maximum attainable value
of the log-likelihood function with respect to the change in constraints. 
If $H_0: h(\theta) = c$ is true and $l(\widetilde{\theta})$ gives the optimal value, $\widetilde{\lambda}$ should be close to zero. Given this ``economic" interpretation in terms of Lagrange multipliers, it is not surprising that econometricians prefer the term LM rather than RS. In terms of Lagrange multipliers, (15) can be expressed as
\begin{equation}
    RS = LM = \widetilde{\lambda}^\prime H(\widetilde{\theta})^\prime \mathcal{I}(\widetilde{\theta})^{-1}H(\widetilde{\theta})\widetilde{\lambda}.
\end{equation}

After \cite{Byron:1968}, it took another decade for econometricians to realize the potential of the RS test. The earlier notable contributions include \cite{Savin:1976}, \cite{Savin:1977}, \citet{Breusch:1978, Breusch:1979} and \citet{Godfrey78a, Godfrey78b, Godfrey78c}. Possibly \cite{Breusch:1980} had been the most influential. They collected relevant research reported in the Statistics literature, presented the RS test in a general framework in the context of evaluating various econometric models, and discussed many applications. In a full length research monograph, \cite{Godfrey:1988} provided a comprehensive account of most of the available RS tests in Econometrics. \cite{Bera:1991} and 
\cite{Bera:2001} demonstrated that many of the commonly used specification tests could be given a score-test interpretation. For the last two-score years the RS tests had been the most common items in econometricians' kit for testing tools. It is not hard to understand the popularity of the score test principle in economics. In most cases, the algebraic forms of W and LR tests can hardly be simplified beyond their original formulae (16) and (17). On the other hand, in the majority of the cases the RS test statistics can explicitly be reduced to neat and elegant explicit formulae enabling its easy incorporation into computer software.

We will not make any attempt to provide a comprehensive list of applications of the RS test in Econometrics, for there are far too many. For instance, consider the workhorse of basic econometric modeling, the linear regression model:
\begin{equation}
    y_i = x_i^\prime \beta + \epsilon_i,
\end{equation}
where $y_i$ is the $i$-th observation on the dependent variable, $x_i$ is the $i$-th observation on $k$ exogenous 
variables and $\epsilon_i \sim IID N(0, \sigma^2), \ i = 1, 2, \ldots, n$. The ordinary least squares (OLS) estimation and the related hypotheses tests are based on the four basic assumptions: correct linear functional form; the assumptions of disturbance normality; 
homoskedasticity;
 and serial independence. Just to name some of the uses of the RS test principle, test for normality was derived by \cite{Bera:1981} and \cite{JB:1987}; \cite{BreuschPagan79} proposed a test for 
 homoskedasticity; and \citet{Godfrey78a, Godfrey78b} developed tests for serial independence which are very close to the earlier \cite{DW1950} test. 

To see the attractiveness of the RS test, let us briefly consider the popular Jarque and Bera (JB) test for normality. \cite{Bera:1981} started with the \cite{Pearson:1895} family of distributions for the disturbance term $\epsilon_i$ in (36). That means if the pdf of $\epsilon_i$ is $f(\epsilon_i)$, we can write 
\begin{equation}
    \frac{d\log f(\epsilon_i)}{d\epsilon_i} = \frac{c_1 - \epsilon_i}{\sigma^2 - c_1\epsilon_i + c_2\epsilon_i^2}, \qquad  i = 1, 2, \ldots, n,
\end{equation}
where $c_1$ and $c_2$ are constants. The null hypothesis of normality can be stated as $H_0: c_1 = c_2 = 0$ in (37). Given the complexity of ML estimation of $\sigma^2$, $c_1,$ and $c_2$ in the Pearson family of distributions, W and LR tests are ruled out from a practical point of view. However, the score functions corresponding to $c_1$ \text{and} $c_2$ in (37), evaluated under the normality assumption, are given respectively by
\begin{equation}
  S(\tilde{c}_1) = \frac{n\sqrt{b_1}}{3}  
\end{equation}
and
\begin{equation}
    S(\tilde{c}_2) = \frac{n}{4}(b_2 - 3),
\end{equation}
where $\sqrt{b_1} = {m_3}/{m^{3/2}_2} $ and $b_2 = {m_4}/{m^2_2}$ with $m_j = \frac{1}{n}\sum^n_{i = 1} \tilde{\epsilon}_i^j$, $\tilde{\epsilon}_i = y_i - x^\prime_i\tilde{\beta}$ as OLS residuals, $j = 2, 3, 4$. For large $n$, under normality
\begin{equation}
    E\left[\sqrt{b_1}\right] = 0, \quad Var\left[\sqrt{n b_1}\right] = 6,
\end{equation}
\begin{equation}
    E\left[b_2\right] = 3, \quad Var\left[\sqrt{n} b_2\right] = 24,
\end{equation}
and they are asymptotically normally distributed. Thus, a simple test statistic for normality is given by
\begin{equation}
    JB = n\left[\frac{(\sqrt{b_1})^2}{6} + \frac{(b_2 - 3)^2}{24}\right],
\end{equation}
which is asymptotically distributed as $\chi^2_2$. It turns out that this test was mentioned by \cite{Bowman:1975} but was hardly used in practice due to its lack of theoretical underpinnings. The RS test principle uncovered the theoretical justification of (42), ensuing the asymptotic optimality of the test. As it is obvious, JB is based on the two moments, third and fourth. One could have started with these two moments directly without going through the full derivations. From that point of view this RS test has a \textit{moment} test interpretation.

It is quite common to express specification tests in Econometrics 
as \textit{moment} tests. In a way ``any" moment test can be obtained as a RS test under a suitably defined density function. To see this, let us write the $r$ moment 
restrictions as 
\begin{equation}
    E_f[m(y; \theta)] = 0,
\end{equation}
where $E_f[\cdot]$ means that (43) is true only when $f(y; \theta)$ is the correct pdf. A test for the hypothesis $H_0: E_f[m(y; \theta)] = 0$ can be based on the estimate of the sample counterpart of $E_f[m(y; \theta)]$, namely, 
\begin{equation}
    \frac{1}{n}\sum^n_{i = 1}m(y_i; \theta).
\end{equation}
Now consider an auxiliary density function
\begin{equation}
    f^*(y; \theta, \gamma) = f(y; \theta)\exp[\gamma^\prime m(y;\theta) - \phi(\theta, \gamma)],
    \label{aux-pdf}
\end{equation}
where $\phi(\theta, \gamma) = \ln\int \exp[\gamma^\prime  m(y; \theta)]f(y;\theta)dy$, with
$\gamma$ as $(r\times 1)$ parameter vector.

Note that if $f(y; \theta)$ is the correct pdf, then $\gamma = 0$ in (\ref{aux-pdf}). The log-likelihood function under the alternative hypothesis is
\begin{equation}
    l^*(\theta, \gamma) = \sum^n_{i = 1}\ln f^*(y_i; \theta, \gamma).
\end{equation}
Therefore, the score function for testing $\gamma = 0$ in (\ref{aux-pdf}) is given by
\begin{equation}
    \left. \frac{\partial l^*(\theta, \gamma)}{\partial \gamma} \right|_{\gamma = 0} = \sum^n_{i = 1}m(y_i, \theta),
\end{equation}
and it provides the identical moment test as in (44). This interpretation of the moment test as a score test was first noted by \cite{White:1994}. It is easy to see that there are many choices of auxiliary pdf $f^*(y; \theta, \gamma)$ and the score test will be invariant with respect to these choices, as depicted in Figuire 1. The LR and W tests, however, will be sensitive to the forms of $f^*(y; \theta, \gamma)$. This 
ends our coverage of the use of the RS test in Econometrics.
 
\section{Some applications of the RS test in Statistics}


\cite{Rao:1950seq} proposed a sequential test of null hypotheses based on the score statistic and his work on locally most powerful (LMP) tests in the case of one-sided alternative hypotheses. His proposal was a reaction to \cite{Wald:1945} sequential probability ratio test (SPRT) which was based on the idea of likelihood ratio test for the fixed sample case. 

Wald's SPRT statistic was devised to discriminate between different alternative hypotheses for the value of the unknown parameter $\theta$. On the other hand, \cite{Rao:1950seq} seeks to test a null hypothesis 
$H_0\!: \theta=\theta_0$  against a one-sided alternative hypothesis 
$H_1\!: \theta>\theta_0$ with a test statistic that depends only on the null value.

For the fixed sample case, with sample of size $N$, the LMP test suggested by 
\cite{RaoPoti:1946} is defined by [also see equation (6)]
\begin{equation}
P_N^\prime(\theta_0) \ge \mu \, P_N(\theta_0), 
\end{equation}
where $P_N^\prime(\theta_0)$ is the first derivative of $P_N(\theta)$ at $\theta=\theta_0$, with $\mu$ chosen so as to maintain Type-I error at a predetermined level. Motivated by this result, \cite{Rao:1950seq} proposes a sequential test of the form
\begin{equation}
	P_n^\prime(\theta_0) \ge A(N) \, P_n(\theta_0), 
	\label{Rao50}
\end{equation}
with $n\le N$, $A(N)$ a properly determined constant depending on the overall level of significance, and $N$ being the upper limit to the number of observations. According to this sequential testing scheme, the sampling stops with rejection of the null hypothesis, at the smallest value of $n$ for which the inequality (\ref{Rao50}) holds true. If by the $N$th sampled unit (\ref{Rao50}) is not realized, the null is not rejected.

\cite{Berk:1975} proved that the sequential score tests against a one-sided alternative, where the stopping rule is the first time a certain random walk exceeds a bounded interval, are LMP tests \textit{asymptotically}.

Sequential testing procedures that perform interim analyses during the evolution of the experiment, with the goal of obtaining the result earlier than the termination time suggested by the fixed sample analysis due to time or monetary cost considerations or ethical reasons, are easier validated with the use of score-based test statistics rather than the analogues of LR statistics. We may refer to chapters 9-11 of \cite{Sen1981book} for the role of score processes in sequential nonparametrics, where it is mentioned (p.339) ``it is comparatively simpler to verify these regularity conditions [i.e., for the score] than those for the likelihood function.''


\cite{Lombardi:1951}, in a thesis on how to select a panel of judges for taste testing and quality evaluation using scientifically sound methods, appears to be one of the first applications of the sequential testing using \cite{Rao:1950seq} methodology and its comparison with Wald's approach.

To diagnose the potential ability of candidate judges and to decide on the selection of a taste panel, each candidate judge is required to perform a prespecified number of sample comparisons.
The number of sample comparisons that should be performed by each candidate judge 
before reaching to a decision on who to include in the taste panel is always a concern.


\cite{Bradley:1953}, in conjunction with Lombardi, adapted Rao's method to binomial distribution,  and is an early effort to communicate these statistical procedures for the selection of a taste panel to food technologists.  

What makes the Rao procedure relatively more appropriate than that of Wald is that a limit to the testing of any one potential judge may be set. In this application, $N$ denotes the maximum number of tests to be given to any judge. As it is noted by \cite[p.28]{Bradley:1953}: 
``The theory of the procedure needs further
investigation since its properties are not well known. However, when
it is applied to sequences of triangle tests, apparently satisfactory
results are obtained."

In another context, time-series researchers use sequential analysis to determine and test for
structural breaks. In a recent application, \cite{Bucci:2024} proposes a sequentially computed score statistic to test for the number of regimes in multivariate nonlinear models.   

\subsection{The role of the score statistic in survival analysis}

Another context where the RS test statistic has found fruitful applications for inference is the analysis of survival data.  The semiparametric proportional hazards model proposed by \cite{Cox:1972} is a standard tool of analysis for time-to-an-event data met in medical, engineering and economic applications.  The parameter estimation using the partial likelihoood of \cite{Cox:1975} initiated an intense research activity for the validation of inference. The majority of the test statistics are special cases of weighted score statistics for different weighting functions and different type of covariates. 

The partial likelihood score statistic has a natural martingale characterization. By rewriting the model within the counting process framework, \cite{AndersenGill:1982} were able to obtain a general asymptotic theory of the score statistic and the associated estimator. In a research related to sequentially computed score test statistic for repeated significance tests, \cite{Tsiatis:1981} established the joint asymptotic normality of efficient scores test for the proportional hazards model calculated over time. In a fundamental breakthrough, \cite{Sellke:1983} showed that the score process (over time) of the partial likelihood is approximated by a suitable martingale and thus behaves asymptotically like Brownian motion.      


\cite{Bilias:2000} offered an application of 
a repeated significance test in a retrospective analysis of the Pennsylvania `Reemployment Bonus'
controlled experiments conducted by the US Department of Labor. 
Their main purpose was to determine whether the offer of a bonus amount to the unemployment insurance
(UI) claimants, provided that they find a job with some required permanence within a given
period of time, can act as an incentive for more intensive job-seeking with subsequent reduction
of the unemployment spells.
The response of primary interest is the \textit{length of insured unemployment
	spell} and it is assumed that it follows a proportional hazard
regression model. The statistic for measuring the effect of the various bonus
packages on the duration of insured unemployment relative to the existing scheme 
is the partial likelihood score statistic. In
carrying out sequential analysis, the score statistic is evaluated repeatedly, at
different points in chronological time, each time with the available data. 
The retrospective sequential analysis concluded that the experiment could be concluded earlier than the fixed sample analysis with gains in time and monetary savings.

\section{Robust RS tests under distributional and parametric misspecifications}

As we have narrated in the previous sections the success of the RS test had been phenomenal. However the main problem in these specification tests is that they are developed under the assumption that the underlying probability model is \textit{correctly specified}. When the assumed model is misspecified, it is well known that the RS test loses its local optimal properties. 

While discussing the problem in statistical hypothesis testing, \citet[pp.65-66]{Haavelmo:1944} stated, ``Whatever be the principles by which we choose a ``best" critical region of size $\alpha$, the essential thing is that a test is always developed with respect to a \textit{given fixed} set of possible \textit{alternatives} $\Omega^0$." Haavelmo called $\Omega^0$, the \textit{a priori admissible hypotheses}  and according to him, a test is not robust if we shift our attention to another admissible set $\Omega^\prime$ (that may be obtained by extending $\Omega^0$ to include new/different alternatives), for which the proposed test has poor size and power properties. 

Very often it is difficult to interpret the results of a test applied to a \textit{misspecified} model. For instance, while testing the significance of some of the regression coefficients in the \textit{linear} regression models, the results are not easily interpretable when a \textit{nonlinear} model is the appropriate one [see, \cite{White:1980IER}, \cite{BeraByron:1983} and \cite{Byron:1983}]. In the Statistics and Econometrics literature, most emphasis has been put on the minimization of type-I and type-II error probabilities. There are, however, only a few works that seriously consider the consequences and suggest remedies of misspecifying the a priori admissible hypothesis -- which can be called the type-III error.

Note that
the model under our a priori admissible hypothesis could be misspecified in a variety of ways. Here we consider only two kinds: distributional and parametric. In the former case, the assumed probability density function differs from the true data generating process (DGP). 
\cite{Kent:1982} and \cite{White:1982} analyzed this case and suggested a modified version of the RS test that involves adjustment of the variance of the score function. In the parametric misspecification case, the dimension of the assumed parameter space does not match with the true one. \cite{Bera:1993} developed a modified RS test that is valid under the \textit{local} parametric misspecification. 

\subsection{Robust RS test under distributional misspecification}

Let the true DGP be described by the unknown density $g(y)$ \text{and} $f(y;\theta)$ be our assumed distribution. The RS test statistic given in (15) is not valid when $g(y)$ \text{and} $f(y; \theta)$ differ. This is because some of the standard results breakdown under distributional misspecification. For instance, consider the information matrix (IM) equality: 
\begin{equation}
    E_f\left[\frac{\partial \ln f(y; \theta)}{\partial \theta} \cdot \frac{\partial \ln f(y; \theta)}{\partial \theta^\prime}\right] = E_f\left[-\frac{\partial^2\ln f(y;\theta)}{\partial\theta\partial\theta^\prime}\right],
\end{equation}
where $E_f[\cdot]$ denotes expectation under $f(y;\theta)$. Let us now define
\begin{equation}
    J(\theta_g) = n  E_g\left[\frac{\partial \ln f(y; \theta)}{\partial \theta} \cdot \frac{\partial \ln f(y; \theta)}{\partial \theta^\prime} \right]
\end{equation}
\begin{equation}
    K(\theta_g) = n  E_g\left[-\frac{\partial^2 \ln f(y;\theta)}{\partial \theta \partial \theta^\prime}\right],
\end{equation}
where $\theta_g$ minimizes the Kullback-Leibler information criterion [see \cite{White:1982}]
\begin{equation}
    \mathcal{I}_{KL} = E_g\left[\ln\frac{g(y)}{f(y;\theta)}\right].
\end{equation}
One can easily see that $J(\theta_g) \neq K(\theta_g)$, in general. 

\textbf{Example 4:} Suppose we take $f(y; \theta) \equiv N(\mu,\sigma^2)$, and let the DGP $g(y)$ satisfy $E_g[y] = \mu$, $E_g[y - \mu]^2 = \sigma^2$, $E_g[y - \mu]^3 = \mu_3$ and $E_g[y - \mu]^4 = \mu_4$. Then it is easy to show that 
\begin{equation}
    J(\theta_g) = \begin{bmatrix}
\frac{1}{\sigma^2} & \frac{\mu_3}{2\sigma^6}   \\
\frac{\mu_3}{2\sigma^6} & \frac{\mu_4}{4\sigma^2}-\frac{1}{4\sigma^4} \\
\end{bmatrix}
\end{equation}
and
\begin{equation}
    K(\theta_g) = \begin{bmatrix}
\frac{1}{\sigma^2} & 0 \\
0 & \frac{1}{2\sigma^2} \\
\end{bmatrix}.
\end{equation}
Hence, $J(\theta_g) = K(\theta_g)$ if and only if $\mu_3 = 0$ and $\mu_4 = 3\sigma^4$. We can clearly see the connection of these conditions and the JB test for normality given in (42). 

Due to this divergence between $J$ and $K$, and noting that we defined the information matrix $\mathcal{I}(\theta)$ in (10) by taking expectation under $f(y; \theta)$ instead of under the DGP $g(y)$, the standard RS test in (15) is not valid.  
Let us define an estimator of $\theta$ by maximizing a likelihood function based on the 
misspecified density $f(y; \theta)$ in place of the unknown DGP $g(y)$.
Such an estimator is called quasi-MLE (QMLE). 
An early reference to QMLE can be found in \citet[p.135]{Koopmans:chap}
[for more on this see \cite{BBY2020}].
We will denote QMLE of $\theta$ (under $H_0$) by $\tilde{\theta}$.
\cite{Kent:1982} and \cite{White:1982} suggested the following robust form of the RS test statistic for testing the $H_0:h(\theta)=c$: 
\begin{equation}
    RS^*(D) = S(\widetilde{\theta})^\prime K(\widetilde{\theta})^{-1}H(\widetilde{\theta})[H(\widetilde{\theta})^\prime B(\widetilde{\theta})H(\widetilde{\theta})]H(\widetilde{\theta})^\prime K(\widetilde{\theta})^{-1}S(\widetilde{\theta}),
\end{equation}
where $H(\theta)=\partial h(\theta)/\partial \theta$, $B(\theta) = K(\theta)^{-1}J(\theta)K(\theta)^{-1}$ 
and the notation $RS^*(D)$ is used to signify \textit{robust} RS test statistic under \textit{distributional} misspecification. 

Under $H_0: h(\theta) = c$, $RS^*(D)$ is asymptotcally distributed as $\chi^2_r$ even under distributional misspecification, that is, when the assumed density $f(y; \theta)$ does not coincide with the true DGP $g(y)$. This approach of finding the asymptotically correct formula for variance has its origin in 
\citet[pp.148-150]{Koopmans:chap}; 
for more on this see \cite{Bera:2021}.
Expression (56) can be simplified if the parameter vector $\theta$ $(p\times 1)$ can be 
partitioned as $\theta=(\gamma^\prime, \psi^\prime)^\prime$ where $\gamma$ and $\psi$ have 
dimensions $m$ and $r$, respectively, $m+r=p$, and we test $H_0: \psi=\psi_*$ (say). 
Let us also partition the score function $S(\theta)$ and $J(\theta)$ [similarly $K(\theta)$]
as
\begin{equation}
	S(\theta) = \frac{\partial l(\theta)}{\partial\theta} = \begin{bmatrix}
		\frac{\partial l(\theta)}{\partial\gamma} \\ \frac{\partial l(\theta)}{\partial\psi} 
	\end{bmatrix}
	= \begin{bmatrix}
		S_\gamma(\theta) \\ S_\psi(\theta) 
	\end{bmatrix}
	\qquad \text{(say)}
\end{equation}
and
\begin{equation}
	J(\theta) =  \begin{bmatrix}
		J_\gamma & J_{\gamma\psi}  \\
		J_{\psi\gamma} & J_{\psi}  
	\end{bmatrix}.
\end{equation}
While testing $H_0: \psi=\psi_*$, under this setup $h(\theta)=\psi - \psi_*$ and $H(\theta)=[0_{r\times (p-r)}, I_{(r\times r)}]$,
and we can express $RS^*(D)$ in (56) as [see also \cite{BBY2020}]
\begin{eqnarray}
	RS_\psi^*(D)
	&=&
	S^\prime_\psi (\widetilde\theta)
	\bigl[
	K_\psi(\widetilde\theta)
	+
	J_{\psi\gamma}(\widetilde\theta)
	J_\gamma^{-1}(\widetilde\theta)K_\gamma(\widetilde\theta)
	J_\gamma^{-1}(\widetilde\theta)J_{\gamma\psi}(\widetilde\theta)
	\nonumber
	\\ [.1in]
	&&
	-
	J_{\psi\gamma}(\widetilde\theta)J_\gamma^{-1}(\widetilde\theta)
	K_{\gamma\psi}(\widetilde\theta)
	-
	K_{\psi\gamma}(\widetilde\theta)
	J_\gamma^{-1}(\widetilde\theta)J_{\gamma\psi}(\widetilde\theta)
	\bigr]^{-1}
	S_\psi (\widetilde\theta),
\end{eqnarray}
where $\tilde{\theta}=(\tilde{\gamma}^\prime, \psi_*^\prime)^\prime$, the restricted MLE under
$H_0: \psi=\psi_*$. 

\textbf{Example 5:} In (37) the parameters $c_1$ and $c_2$ of the Pearson family of distributions can be treated, respectively, as the ``skewness" and ``kurtosis" parameters.
 Suppose we test the symmetry ignoring the (excess) kurtosis. Then we can start with the system (37) with $c_2 = 0$, that is, 
\begin{equation}
    \frac{d \log f(\epsilon_i)}{d\epsilon_i} = \frac{c_1 - \epsilon_i}{\sigma^2 - c_1\epsilon_i}.
\end{equation}
After some derivation, it can be shown that the standard RS test for $c_1 = 0$ is given by 
\begin{equation}
    RS_{c_1} = n\frac{(\sqrt{b_1})^2}{6},
\end{equation}
which is essentially the first part of JB in (42). If $f(\epsilon_i)$ in (60) is not the true DGP, $RS_{c_1}$ will not be valid; in particular, the asymptotic variance formula used in (61), $Var(\sqrt{n}b_1) = 6$ is incorrect [see also equation (40)]. For instance in the presence of excess kurtosis, there will be proportionately more outliers, resulting in higher variance, and thus ``6" will underestimate the true variance of $\sqrt{n}b_1$. After incorporating the variance correction as in $RS^*_\psi(D)$ in (59) the robust RS test statistic can be written as [for further details see \cite{Bera:2017}]: 
\begin{equation}
    RS^*_{c_1}(D) = n\frac{(\sqrt{b_1})^2}{[9 + m_6m^{-3}_2 - 6m_4m^{-2}_2]},
\end{equation}
where $m_j = \frac{1}{n}\sum^n_{i = 1}\tilde{\epsilon}_i^j$, $j = 2, 4, 6$. From (62) we can write the population counterpart of the $Var(\sqrt{n}b_1)$ as 
\begin{equation}
    Var(\sqrt{n}b_1) = 9 + \mu_6\mu^{-3}_2 - 6\mu_4\mu^{-2}_2,
\end{equation}
where $\mu_j$ denotes the $j$-th population moment of $\epsilon$. Therefore, the construction of the robust RS test statistic $RS^*_{c_1}(D)$ indicates that the true variance of $\sqrt{n}b_1$ that is valid under excess kurtosis is given by (63). If we impose normality, $\mu_6 = 15\sigma^6$ and $\mu_4 = 3\sigma^4$, then with $\mu_2 = \sigma^2$, (63) reduces to $Var(\sqrt{n}b_1) = 9 + 15 - 6 \times 3 = 6$, as in (61). 

\textbf{Example 6:} Consider the test for 
homoskedasticity under the regression framework of (36), where we now explicitly specify the 
heteroskedastic structure 
as $Var(\epsilon_i) = \sigma^2_i = \sigma^2 + \delta^\prime z_i$, where $\delta$ is a $r \times 1$ vector and $z_i$'s are fixed exogenous variables, $i = 1, 2, \ldots ,n$. Assuming \textit{normality} of $\epsilon_i$ the RS statistic for testing 
homoskedasticity hypothesis $H_0: \delta = 0$ is given by [see \cite{BreuschPagan79}]
\begin{equation}
    RS_\delta = \frac{\nu^\prime Z(Z^\prime Z)^{-1} Z^\prime \nu}{2\widetilde{\sigma}^4},
\end{equation}
where $\nu_i = \widetilde{\epsilon}^2_i - \widetilde{\sigma}^2$, $\nu = (\nu_1, \nu_2, \ldots, \nu_n)^\prime$ and $Z = (z_1, z_2, \ldots, z_n)^\prime$. The factor ``$2\widetilde{\sigma}^4$" is the consequence of the normality assumption, and therefore, the test in (64) will not be valid even asymptotically if $\epsilon_i$'s are not distributed as normal. Using (61), the robust form of RS test statistics can be derived as 
\begin{equation}
    RS^*_\delta(D) = \frac{\nu^\prime Z(Z^\prime Z)^{-1}Z^\prime \nu}{\frac{\nu^\prime \nu}{n}}.
\end{equation}
This is the same modification suggested by \cite{Koenker:1981}. Note that the modification amounts to replacing $Var(\epsilon^2_i) = \mu_4 - \mu^2_2 = 3\sigma^4 - \sigma^4 = 2\sigma^4$ (derived under normality) by a robust estimate, namely, by, $\frac{1}{n}\sum^n_{i = 1}(\widetilde{\epsilon}_i^2 - \widetilde{\sigma}^2)^2 = ({\nu^\prime \nu})/{n}$. For other applications of $RS^*(D)$ see for instance, \cite{Lucas:1998} and \cite{Bera:2017}.
\par
In a similar fashion the Wald statistic in (16) can be robustified as [see \cite{Kent:1982}, \cite{White:1982}, and \cite{PaceSalvan:1997}]: 
\begin{equation}
    W^* = [h(\hat{\theta})-c]^\prime[H(\hat{\theta})B(\hat{\theta})H(\hat{\theta})]^{-1}[h(\hat{\theta}) - c],
\end{equation}
and asymptotically it has $\chi^2_r$ distribution under the null hypothesis $H_0: h(\theta) = c$. Thus, robust $RS^*$ and $W^*$ are obtained by robustifying the variance expressions, respectively, of $S(\widetilde{\theta})$ \text{and} $h(\hat{\theta})$. However, similar robustification of LR statistic in (17) is not possible. \cite{Kent:1982} showed that under distributional misspecification LR statistic is asymptotically distributed as a weighted sum of $r$ independent $\chi^2_1$ variables, and thus no obvious ``variance" adjustment is possible.

\subsection{Robust RS tests under parametric misspecification}

Consider a general statistical model represented by the log-likelihood function $l(\gamma, \psi, \phi)$ where $\gamma, \psi, \text{and }\phi$ are parameter vectors with dimensions $(m \times 1)$, $(r \times 1)$ \text{and} $(q \times 1)$, respectively. Thus our $(p\times 1)$ parameter vector is $\theta = (\gamma^\prime, \psi^\prime, \phi^\prime)^\prime$ and $p = m + r + q$. Suppose an investigator sets $\phi = 0$ and tests $H_0: \psi = 0$ using the log-likelihood function $l_1(\gamma, \psi) = l(\gamma, \psi, 0)$. We will denote the RS statistic for testing $H_0$ in $l_1(\gamma, \psi)$ by $RS_\psi$. Let us also denote $\widetilde{\theta} = (\widetilde{\gamma}^\prime, 0, 0)^\prime$, where $\widetilde{\gamma}$ is MLE of $\gamma$ when $\psi = 0$ \text{and} $\phi = 0$. The score vector and the information matrix are defined, respectively, as 
\begin{equation}
    S(\theta) = \frac{\partial l(\theta)}{\partial\theta} = \begin{bmatrix}
\frac{\partial l(\theta)}{\partial\gamma} \\ \frac{\partial l(\theta)}{\partial\psi} \\ \frac{\partial l(\theta)}{\partial\phi} \\
\end{bmatrix}
 = \begin{bmatrix}
S_\gamma(\theta) \\ S_\psi(\theta) \\ S_\phi(\theta) \\
\end{bmatrix}
\qquad \text{(say)}
\end{equation}
\begin{equation}
    \mathcal{I}(\theta) = E_\theta\left[-\frac{\partial^2l(\theta)}{\partial\theta\partial\theta^\prime}\right] = \begin{bmatrix}
\mathcal{I}_\gamma & \mathcal{I}_{\gamma\psi} & \mathcal{I}_{\gamma\phi} \\
\mathcal{I}_{\psi\gamma} & \mathcal{I}_{\psi} & \mathcal{I}_{\psi\phi} \\
\mathcal{I}_{\phi\gamma} & \mathcal{I}_{\phi\psi} & \mathcal{I}_{\phi} \\
\end{bmatrix}.
\end{equation}
If $l_1(\gamma, \psi)$ were correctly specified, then the RS test statistic of (15), in the current context can be written as 
\begin{equation}
    RS_\psi = S_\psi(\widetilde{\theta})^\prime\mathcal{I}_{\psi\cdot\gamma}^{-1}(\widetilde{\theta})
    S_\psi(\widetilde{\theta}),
\end{equation}
where $\mathcal{I}_{\psi\cdot\gamma} = \mathcal{I}_\psi - \mathcal{I}_{\psi\gamma}\mathcal{I}_{\gamma}^{-1}\mathcal{I}_{\gamma\psi}$ and it will be asymptotically distributed as \textit{central} $\chi^2_r$. Under this set-up, asymptotically $RS_\psi$ will have the correct size and will be locally optimal. 
\par
Let us now consider the case of \textit{parametric misspecification}. Suppose the true log-likelihood function is $l_2 = (\gamma, \phi) = l(\gamma, 0, \phi)$, so that the alternative $l_1(\gamma, \psi)$ becomes misspecified. Using the sequence of \textit{local} DGP $\phi = {\delta}/{\sqrt{n}}$, \cite{Davidson:1987} and \cite{Saikkonen:1989} showed that under $l_2(\gamma, \phi)$ with $\phi = {\delta}/{\sqrt{n}}$, $RS_\psi$ in (69), under $H_0: \psi = 0$ will be distributed as \textit{non-central} $\chi^2_r$ with noncentrality parameter, 
\begin{equation}
    \lambda(\delta) = \delta^\prime\mathcal{I}_{\phi \psi \cdot \gamma}\mathcal{I}^{-1}_{\psi \cdot \gamma} \mathcal{I}_{\psi \phi \cdot \gamma}\delta,
\end{equation}
with $\mathcal{I}^\prime_{\phi\psi \cdot \gamma} = \mathcal{I}_{\psi\phi \cdot \gamma} = \mathcal{I}_{\psi\phi} - \mathcal{I}_{\psi\gamma}\mathcal{I}^{-1}_\gamma \mathcal{I}_{\gamma\phi}$. Owing to the presence of this non-centrality parameter, $RS_\psi$ will reject the null hypothesis $H_0: \psi = 0$
more often than allowed by the preassigned size of the test, even when $\psi=0$. Therefore, under parametric misspecification, $RS_\psi$ will have an excessive size. For the expression of $\lambda(\delta)$ \text{in} (70), we note that the crucial quantity is $\mathcal{I}_{\psi\phi \cdot \gamma}$, which can be interpreted as the conditional covariance between the scores $S_\psi$ \text{and} $S_\phi$ given $S_\gamma$. If $\mathcal{I}_{\psi\phi \cdot \gamma} = 0$, then the local presence of the misspecified parameter $\phi = {\delta}/{\sqrt{n}}$ will have no effect on the performance of $RS_\psi$.
\par
Using the expression in (70), \cite{Bera:1993} suggested a modification to $RS_\psi$ so that the resulting test is robust to the presence of $\phi$. The modified statistic is given by
\begin{equation}
    \begin{split}
        RS^*_\psi(P) = & [S_\psi(\widetilde{\theta}) - \mathcal{I}_{\psi\phi \cdot \gamma}(\widetilde{\theta})\mathcal{I}^{-1}_{\phi \cdot \gamma}(\widetilde{\theta})S_\phi(\widetilde{\theta})]^\prime \\
        & [\mathcal{I}_{\psi \cdot \gamma}(\widetilde{\theta}) - \mathcal{I}_{\psi\phi \cdot \gamma}(\widetilde{\theta})\mathcal{I}^{-1}_{\phi \cdot \gamma}(\widetilde{\theta})\mathcal{I}_{\phi\psi \cdot \gamma}(\widetilde{\theta})]^{-1} \\
        & [S_\psi(\widetilde{\theta}) - \mathcal{I}_{\psi\phi \cdot \gamma}(\widetilde{\theta})\mathcal{I}^{-1}_{\phi \cdot \gamma}(\widetilde{\theta})S_\phi(\widetilde{\theta})].
    \end{split}
\end{equation}
Here the notation $RS^*(P)$ is used to signify \textit{robust} RS test statistic under \textit{parametetric} misspecification. Under $H_0: \psi = 0$, $RS^*_\psi(P)$ is asymptotically distributed as \textit{central} $\chi^2_r$, i.e., $RS^*_\psi(P)$ has the \textit{same} asymptotic distribution as of $RS_\psi$ in (69) based on the correct specification. 
Thus, $ RS^*_\psi(P)$ provides an asymptotically correct-size test under the locally misspecified alternative $l_2(\gamma, \phi) $. 
\par
$RS^*_\psi(P)$ essentially adjusts the asymptotic mean and variance of standard (unadjusted) $RS_\psi$. Another way to look at $RS^*_\psi(P)$ is to view the quantity, $\mathcal{I}_{\psi\phi \cdot \gamma}(\widetilde{\theta})\mathcal{I}_{\phi \cdot \gamma}(\widetilde{\theta})^{-1}S_\phi(\widetilde{\theta})$ as the prediction of $S_\psi(\widetilde{\theta})$ \text{by} $S_\phi(\widetilde{\theta})$. Here $S_\phi(\widetilde{\theta})$ is the score function of the parameter vector $\phi$ whose effect we want to take into account in constructing the robust version of the test. Therefore,
the net score
 $S^*_\psi(\widetilde{\theta})=S_\psi(\widetilde{\theta}) - \mathcal{I}_{\psi\phi \cdot \gamma}(\widetilde{\theta})\mathcal{I}^{-1}_{\phi \cdot \gamma}(\widetilde{\theta})S_\phi(\widetilde{\theta})$ is the part of $S_\psi(\widetilde{\theta})$ that remains after eliminating the effect of $S_\phi(\widetilde{\theta})$. In summary, $S^*_\psi(\widetilde{\theta}\perp S_\phi(\widetilde{\theta})$, though $S_\phi(\widetilde{\theta})$ has ``peer" effect on 
 $S_\psi(\widetilde{\theta})$.
Three more things regarding $RS^*_\psi(P)$ are worth noting. \textit{First}, $RS^*_\psi(P)$ requires estimation only under the joint null, namely for the constrained model in which both $\psi = 0$ and $\phi = 0$. Given the full specification of the model $l(\gamma, \psi, \phi)$, it is of course possible to derive a RS test for $H_0: \psi = 0$ in the presence of $\phi$. However, that requires the MLE of $\phi$, which could be difficult to obtain in some cases. \textit{Second}, when $\mathcal{I}_{\psi\phi \cdot \gamma} = 0$, $RS^*_\psi(P) = RS_\psi$. This is a simple condition to check in practice. As mentioned earlier, if this condition is true, $RS_\psi$ is an asymptotically valid test in the local presence of $\phi$. \textit{Finally}, \cite{Bera:1993} showed that for local misspecification $RS^*_\psi(P)$ is asymptotically equivalent to \cite{Neyman:1959} $C(\alpha)$ test, and therefore, shares its optimality properties. 

\textbf{Example 7:} To illustrate the usefulness of the robust 
score statistic $RS^*_\psi(P)$, we now consider the tests developed in 
\cite{Anselin:1996} for the mixed regressive - spatial autoregressive (SAR) model 
with a SAR disturbance
\begin{equation}
    \begin{split}
        y &= \phi Wy + X\gamma + u \\
        u &= \psi W u + \epsilon \\
        \epsilon & \sim N(0, I\sigma^2).
    \end{split}
\end{equation}
In this model, $y$ is an $(n \times 1)$ vector of observations on a dependent variable recorded at each of $n$ locations, $X$ is an $(n \times m)$ matrix of exogenous variables, and $\gamma$ is a $(m \times 1)$ vector of parameters, $\phi$ \text{and} $\psi$ are scalar spatial parameters and $W$ is an observable spatial weight matrix with positive elements, associated with the spatially lagged dependent variable and SAR disturbance $u$. This spatial weight matrix represents ``degree of potential interactions" among neighboring locations and are scaled so that the sum of the each row elements of $W$ is equal to one. 
\par
The conventional RS statistic for testing $H_0: \psi = 0$ is given by
\begin{equation}
    RS_\psi = \frac{\left[{\widetilde{u}^\prime W \widetilde{u}}/{\widetilde{\sigma}^2}\right]^2}{T},
\end{equation}
where $\widetilde{u} = y - X\widetilde{\gamma}$ are the OLS residuals, $\widetilde{\sigma}^2 = {\widetilde{u}^\prime \widetilde{u}}/{n}$ and $T = tr[(W^\prime + W)W]$. One very interesting observation here is that $RS_\psi$ is essentially same as the widely used \cite{Moran:1948} \textbf{I} test. Let us now consider testing $H_0$ under the local presence of $\phi$. First, the crucial quantity to consider is $\mathcal{I}_{\psi\phi \cdot \gamma}$ which is equal to $T$ and that can never be zero. Therefore robustification of $RS_\psi$ is needed. 
\cite{Anselin:1996} derived the robust test as 
\begin{equation}
    RS^*_\psi(P) = \frac{\left[{(\widetilde{u}^\prime W\widetilde{u})}/{\widetilde{\sigma}^2} - T({\mathcal{I}}_{\phi \cdot \gamma})^{-1}{(\widetilde{u}^\prime Wy)}/{\widetilde{\sigma}^2}\right]^2}{T[1 - T(\mathcal{I}_{\phi \cdot \gamma})^{-1}]},
\end{equation}
where
\begin{equation}
    \mathcal{I}_{\phi \cdot \gamma} = \frac{[(WX\widetilde{\gamma})^\prime M(WX\widetilde{\gamma}) + T\widetilde{\sigma}^2]}{\widetilde{\sigma}^2},
\end{equation}
with $M = I- X(X^\prime X)^{-1}X^\prime$. A comparison of (73) and (74) clearly reveals that $RS^*_\psi(P)$ modifies the standard $RS_\psi$ by correcting the asymptotic mean and variance of the score function $S_\psi$. 
\par

In a similar way we can find $RS_\phi$ and $RS^*_\phi(P)$ which are given, respectively, by  
\begin{equation}
    RS_\phi = \frac{\left[{(\widetilde{u}^\prime W y)}/{\widetilde{\sigma}^2}\right]^2}{\mathcal{I}_{\phi \cdot \gamma}}
\end{equation}
and 
\begin{equation}
    RS^*_\phi (P) = \frac{\left[{(\widetilde{u}^\prime W y)}/{\widetilde{\sigma}^2} - {(\widetilde{u}^\prime W \widetilde{u})}/{\widetilde{\sigma}^2}\right]^2}{\mathcal{I}_{\phi \cdot \gamma} - T},
\end{equation}
where
 $\mathcal{I}_{\phi\cdot\gamma} = \mathcal{I}_\phi - \mathcal{I}_{\phi\gamma}\mathcal{I}_{\gamma}^{-1}\mathcal{I}_{\gamma\phi}$ using 
the submatrices of the partinioned form of $\mathcal{I}(\theta)$ given in (68).
\cite{Anselin:1988} derived a joint RS test for $H_0: \psi = \phi = 0$ under the framework of (72) and that takes the following form
\begin{equation}
    RS_{\psi\phi} = \frac{\left[{(\widetilde{u}^\prime W \widetilde{u})}/{\widetilde{\sigma}^2}\right]^2}{T} + \frac{\left[{(\widetilde{u}^\prime Wy)}/{\widetilde{\sigma}^2} - {(\widetilde{u}^\prime W \widetilde{u})}/{\widetilde{\sigma}^2}\right]^2}{\mathcal{I}_{\phi \cdot \gamma} - T}.
\end{equation}
This statistic is asymptotically distributed $\chi^2_2$. It is easy to verify that
[see \citet[Corollary 1]{BBY2020}] 
\begin{equation}
    RS_{\psi\phi} = RS_\psi + RS^*_\phi(P) = RS_\phi + RS^*_\psi(P).
\end{equation}
In other words, the directional RS test for $\psi$ \text{and} $\phi$ 
can be decomposed into sum of 
the unadjusted one-directional test for one type of alternative and the 
adjusted form for the other alternative. Equalities in (79) can facilitate computations of the adjusted (robust) RS tests after having the unadjusted versions which are easy to obtain and are reported in most of the spatial software.

\cite{Anselin:chap} and \cite{Anselin:1996} provided simulation results on the finite sample performance of the unadjusted and adjusted RS tests and some 
related 
tests. The adjusted tests $RS^*_\psi(P)$ and $RS^*_\phi(P)$ performed remarkably well. Those had very reasonable empirical sizes, remaining within 
the confidence intervals in all cases. 
In terms of power they performed exactly the way they were supposed to. 

\subsection{Robust RS tests under \textit{both} the distributional and parametric misspecifi-\\\hspace*{-1.27cm}cations}

Now we combine the results of Sections 5.1 and 5.2 and develop robust tests $RS^*_\psi (DP)$
which provides a two-way protection against both types of misspecifications, distributional 
$(D)$ and parametric $(P)$. As we have noted in (59), $RS^*_\psi (D)$ involves both the 
$J(\theta)$ and $K(\theta)$ matrices in the variance expression of $S_\psi(\tilde{\theta})$.
While to account of the parametric misspecification, as we did in (71), 
$\mathcal{I}_{\psi\phi \cdot \gamma}(\widetilde{\theta})\mathcal{I}^{-1}_{\phi \cdot \gamma}(\widetilde{\theta})S_\phi(\widetilde{\theta})$ must be subtracted
from $S_\psi(\tilde{\theta})$ to center its mean to zero.
The expression for $RS^*_\psi (DP)$ is given by [for details see \cite{BBY2020}]:
\begin{eqnarray}
	RS_\psi^*(DP)&=&
	\bigl[
	S_\psi(\widetilde \theta)
	- J_{\psi\phi\cdot\gamma}(\widetilde \theta)
	J_{\phi\cdot\gamma}^{-1}(\widetilde \theta)
	S_\phi (\widetilde \theta)
	\bigr]^\prime
	\nonumber \\[.1in]
	&&
	\quad \bigl[
	B_{\psi\cdot\gamma}(\widetilde \theta)
	+
	J_{\psi\phi\cdot\gamma}(\widetilde \theta)
	J_{\phi\cdot\gamma}^{-1}(\widetilde \theta)
	B_{\phi\cdot\gamma}(\widetilde \theta)
	J_{\phi\cdot\gamma}^{-1}(\widetilde \theta)
	J_{\phi\psi\cdot\gamma}(\widetilde \theta)
	\nonumber \\[.1in]
	&&
	\quad -
	J_{\psi\phi\cdot\gamma}(\widetilde \theta)
	J_{\phi\cdot\gamma}^{-1}(\widetilde \theta)
	B_{\phi\psi\cdot\gamma}(\widetilde \theta)
	B_{\psi\phi\cdot\gamma}(\widetilde \theta)
	J_{\phi\cdot\gamma}^{-1}(\widetilde \theta)
	J_{\phi\psi\cdot\gamma}(\widetilde \theta)
	\bigr]^{-1}
	\nonumber \\[.1in]
	&& \quad \bigl[ S_\psi(\widetilde \theta) -
	J_{\psi\phi\cdot\gamma}(\widetilde \theta)
	J_{\phi\cdot\gamma}^{-1}(\widetilde \theta) S_\phi (\widetilde
	\theta) \bigr],
\end{eqnarray}
where
\begin{equation}
B_{\psi\cdot\gamma}= K_\psi + J_{\psi\gamma}J_\gamma^{-1}K_\gamma
J_\gamma^{-1}J_{\gamma\psi}
-
J_{\psi\gamma}J_\gamma^{-1}K_{\gamma\psi}
-
K_{\psi\gamma}J_\gamma^{-1}J_{\gamma\psi},
\end{equation}
similarly $B_{\phi\cdot\gamma}$
and 
\begin{equation}
B_{\psi\phi\cdot\gamma}
=
K_{\psi\phi}
-
J_{\psi\gamma}J_\gamma^{-1}K_{\gamma\phi}
-
K_{\psi\gamma}J_\gamma^{-1}J_{\gamma\phi}
+
J_{\psi\gamma}J_\gamma^{-1}K_\gamma
J_\gamma^{-1}J_{\gamma\phi},
\end{equation}
and similarly $B_{\phi\psi\cdot\gamma}$.  Expressions for the general forms of $J(\theta)$
and $K(\theta)$ are given in (51) and (52) and here we are using their partitioned forms 
for $\theta=(\gamma^\prime, \psi^\prime, \phi^\prime)^\prime$.
Under $H_0:\psi=0$, the $RS^*_\psi(DP)$ test statistic will be asymptotically distributed
as $\chi^2_r$ in the presence of both distributional and parametric misspecifications.
Although $RS^*_\psi(DP)$ has rather a lengthy expression as in (80), it is actually easy 
to compute requiring only $\widetilde{\theta}=(\widetilde{\gamma}^\prime, 0^\prime, 0^\prime)^\prime$.  It can be easily seen that 
under no distributional misspecification, i.e., when $f(y; \theta)\equiv g(y)$,
resulting in $K(\widetilde{\theta})=J(\widetilde{\theta})$,
\begin{equation}
	RS^*_\psi(DP)=RS^*_\psi(P),
\end{equation}
and similarly 
under no parametric misspecification, i.e., 
when $\delta=0$ in $\phi=\delta/\sqrt{n}$, 
\begin{equation}
	RS^*_\psi(DP)=RS^*_\psi(D).
\end{equation}
Finally, trivially when $K=J$ and $\delta=0$, 
\begin{equation}
	RS^*_\psi(DP)=RS_\psi
\end{equation}
as given in (69).

\textbf{Example 8:} Let us briefly go back to Example 7 and now introduce distributional 
misspecification along with the presence of parametric misspecification.
This case has been rigorously considered by \cite{Fang:2014} and they demostrated 
both analytically and through extensive simulations that $RS^*_\psi(P)$ and $RS^*_\phi(P)$ 
as given, respectively in (74) and (77) are valid under non-normality. Therefore,
$RS^*_\psi(DP)=RS^*_\psi(P)$ and $RS^*_\phi(DP)=RS^*_\phi(P)$.
This is a somewhat unsual situation.
For this model as given in (72), information matrix equality does not hold, i.e., 
$J(\theta_g)\neq K(\theta_g)$ [see equations (51)-(53)]. However, still $J^{-1}K
J^{-1}=J^{-1}$. 
This is a serendipitous situation, since no additional adjustment is needed for the 
distributional misspecification. The intuition behind this serendipity is that the 
hypotheses $\psi=0$ and $\phi=0$ relate to the conditional \textit{mean} (first moment) of 
$y$ in (72) (conditional on the neighborhood as captured by the $W$ matrix). However, in general, 
only tests for variance (second moment) and higher moments get affected by non-normality.
A similar case appeared in \cite{BBY2020} where they considered testing for random
effects and serial correlation within an error component model. 
Extensive simulation results are also given in \citet{Koley2022, Koley2024} demonstrating 
the robustness of the RS tests under non-normality in finite sample in 
\textit{spatial regression} model
set up.

\section{Epilogue}

We started this survey paper by stating that C.R.\ Rao's work was always inspired by 
some practical problems.  In his 2003 \textit{Econometric Theory (ET)} Interview 
[see \citet[p.349]{Bera2003ET}],
on the RS test, Rao had the following to say, ``The test evolved in a natural way while 
I was analyzing 
  some genetic data. As I recall, the problem was the 
  estimation of a linkage parameter using data sets from different experiments 
  designed in such a way that each data set had information on the same linkage
  parameter. It was, however, necessary to test whether such an assumption could
  be made because of unforeseen factors affecting the experiments. This required
  a test for consistency of estimates derived from different experimental data sets."
Thus we had a new statistical test principle, after LR and W, motivated by a practical problem
in \textit{genetics}. 
However,  as we have narrated here, the resulting RS test principle has a far reaching 
influence even beyond the Statistics, and in particular, it has become 
one of the most useful model misspecification testing tools
in the Econometrics literature. 
In the history of any scientific field, once in a while there comes a moment for a major 
breakthrough.
Appearance of \cite{Rao:1948}  was such a historical moment. In fact, after that we have not
witnessed any new test principle, beyond the trinity, LR, W and RS.

To keep our exposition simple and to be close to the spirit of \cite{Rao:1948}, we have sticked
to the \textit{likelihood framework}.  However, it is easy to extend the RS test and its various 
ramifications to the generalized method of moments (GMM) and estimating functions (EF) 
frameworks [for more on these, see for instance \cite{Basawa:1991} and \cite{Bera:2010}].
We have also largely confined ourselves to the asymptotic properties and distributions of
the tests. 
However there is a huge literature on the investigation of the finite sample performance
of the RS, particularly, in relation to that of LR and W and finding bootstrap critical values;
for example see, \citet{Mukerjee:1990, Mukerjee93:chap} and \cite{Horowitz:1997}. 

To conclude, we can only speculate what is stored in the future.
Given the current vastness of the field we have lost ``sharp moments of birth", like that of 
\cite{Rao:1948}.
However, considering that 75 years have already been passed, it might be a time for a brand new equally 
good test principle.

{\parindent0pt

}


\section*{Acknowledgements}
We are profoundly thankful to an anonymous referee for her/his careful reading and 
offering many pertinent comments which led to improvement of the paper. 
An earlier version of the paper was presented at the Invited Memorial Session for 
Professor C.\ R.\ Rao,
Joint Statistical Meetings (JSM), Portland, Oregon, August 3-8, 2024. We are thankful to the participants of that conference for their comments, especially to the organizer of the Session, Professor 
Ronald L. Wasserstein, Executive Director of the American Statistical Association (ASA), for giving us the opportunity to present our paper.
This paper was also presented at the Department of Statistics, University of Illinois at Urbana-Champaign (UIUC). We would like to thank the attendees of the UIUC seminar for their constructive feedback that further helped in producing an improved version.
We are grateful to our research assistant (RA) Anirudh Adhikary. 
Without his assistance this paper wouldn't have taken off.
We are also grateful to our other RAs, Rong Yuwen, Tiancheng Guo, Scarlett He, 
Yice Zhang and Wenqi Zeng for their diligent work on compiling the citation 
numbers and a very careful analysis. 
We thank Professor Osman Do\u{g}an for reading 
an earlier version of the paper with great care and thoroughness, and providing 
many pertinent comments with detail suggestions for improvements.  We wish 
we could incorporate all his suggestions. Thanks are also due to Dr.\ Malabika Koley 
for comments that improved the exposition of the paper. At the publication stage, 
the Chair Editor, Professor V.\ K.\ Gupta paid painstaking attention to the 
finest details. Indeed, his suggestions, we believe, greatly improved
the final presentation. We, however, retain the responsibility of any remaining errors.
%
%
%
%
Part of this work was completed when the second author visited 
UIUC.
\vskip0.3cm
\bibliographystyle{sa}
\bibliography{ref}

\end{document}